On the Evolution of Subjective Experience

Jerome A. Feldman (ICSI and UC Berkeley)

*This 2022 revision is largely unchanged from the previous arXiv version, but includes newer references and some additional discussion in italics.* An updated summary of this work has been published as: Feldman, J. (2022). Computation, perception, and mind. *Behavioral and Brain Sciences, 45*, E48. doi:10.1017/S0140525X21001886
A more readable, open access, version is: https://escholarship.org/uc/item/6cs78450
*The companion arXiv article, Mysteries of Visual Experience*
http://arxiv.org/abs/1604.08612   *has also been updated.*

## 1. Prologue

"The most beautiful thing we can experience is the **mysterious**. It is the source of all art and science." Einstein

The ancient *mind-body problem* continues to be one of deepest mysteries of science and of the human spirit. *A compelling contemporary version is expressed by the distinguished physicist Frank Wilczek, (A beautiful Question, p.480).*

*I am, and you are, a collection of quarks, gluons, electrons and photons. I am, and you are, a thinking person.*

To experience some of the mystery of vision, look up from this text and consider the richness of your surrounds. All of this magnificence is produced for your personal experience in ways that remain unknown. For one thing, your eyes shift about two hundred times per minute without your mind noticing it. The resulting mental "illusion" is a large detailed stable image of the scene. The stable image in the mind enables us to act effectively as if our vision system were like a camera. But despite millennia of effort, no one understands the link between the visual system (body) and the mind.

One underlying issue is that terms like "mind" and "consciousness" have many meanings that confound scientific study. The core of the mind-body mystery (often called the hard problem) has always rested on our phenomenal feelings of Subjective Experience (SE) and this term will be used here. An alternative framing is to talk about the *naturalization* of the mind. We now know more than ever about the structure and function of the body and brain, but there is no comparable science of SE in an embodied mind. *An excellent recent book on the historical and ongoing search for an explanation of the mind is (99).*

The overall aim of this article is to shed light on the enduring mystery of human SE. If SE has a conventional embodied substrate, there should be a related evolutionary history. Even though the mind-body problem currently remains a mystery, we can try to localize where reductionist explanations break down. This involves applying computational analysis to the problem of mapping mental phenomena to embodied neural structure.

The philosopher David Chalmers, who has thought as much as anyone about the hard problem (1) says:
"I do not claim that idealism is plausible. No position on the mind–body problem is plausible. Materialism is implausible. Dualism is implausible. Idealism is implausible. Neutral monism is implausible. None-of-the-above is implausible. "

The formulation of the hard problem remains a central issue in contemporary philosophy. A recent book by Weisberg (2) presents a good overview of the main approaches. A core issue for both science and philosophy concerns what SE (or Consciousness) contributes to evolutionary fitness, and this is the focus of the current paper. The latest book on the mind by the philosopher Dan Dennett (3) explores the scientific basis of mental evolution, but focuses on language-based uniquely human capabilities and does not directly address SE.

There is an important difference between the philosophical and scientific framing of the hard problem of the mind. Historically, philosophers made important empirical as well as conceptual contributions. However, with the rise of positivism around 1900, philosophers focused on analyzing words and arguments, and avoided empirical topics. Much of current academic philosophy explicitly precludes science.
 *"For many, philosophy is essentially the a priori analysis of concepts, which can and should be done without leaving the proverbial armchair.*
"https://plato.stanford.edu/entries/concepts/#AttConAna  Section 5.

This stance authorizes the detailed examination of concepts such as panpsychism (everything has some mind) and materialism (mind is an illusion). It also includes extensive treatment of philosophical "zombies" - creatures that have all of the features and capabilities of humans except subjective experience. The Routledge Handbook of Philosophy of Animal Minds (4) explores many other aspects of animal minds from a philosophical perspective. The philosophical approach almost exclusively addresses *Consciousness*, which is much broader than SE and, in fact, is not well defined. A current overview of this work can be found in the 54 chapters of the Blackwell Companion to Consciousness (5). The introduction allows that:
"Consciousness studies are frequently criticized for failing to define precisely what Consciousness is. In this respect there has been little change over the last two centuries"

The alternative expression, Subjective Experience (SE), is much better defined and is generally agreed to be an essential feature of Consciousness. The defining problem of SE is the embodied foundation of your private 1st person feelings. No other initial assumptions are appropriate. Focused SE research can provide important insights on the general problem of Consciousness. It can be studied in (some) non-human animals (93) and avoids many other deep and important aspects of mind. There is good evidence for SE in creatures where there is no indication of many other mental traits such as the sense of self, agency, Narrative Consciousness, language, Higher Order Thought, etc.   We can lump all these traits as meta-cognition (C2 in ref 6) and as being beyond the scope of this article. SE is ubiquitous; everything that we are or can be aware of is an SE, pretty much by definition. Some causes of SE, such as empathy, subliminal input, or dreams, are not usually considered conscious, but do evoke subjective emotion and perception (7).

Scientific approaches to SE, like this one, focus on the standard questions of structure, behavior, and experiment. The goal of science is to explain phenomena. One fairly new development is the realization that the world is not deterministic and that probabilistic formulations are required. There has been remarkable progress in understanding the neural and bodily foundation of much of human behavior and its application to social problems. Sapolsky's recent book "Behave" (40) is an excellent and approachable overview of these advances. A major exception remains where the defining phenomenon is 1st person experience. Unless you believe that you have subjective experience, you can not address the mind-brain problem. The basic problem remains the same even if you try to explain only your own SE.

There is an extensive literature on subjective experience, but almost all of it focuses on two domains – perception and emotions. In fact, SE is important in a wide range of phenomena including speech, dreams, empathy, phantom limbs, tool embodiment, etc. Other SE include time travel, replay, imagination and simulation. The last two terms are often used as synonyms (85). I prefer simulation because it suggests mechanisms. The general term *simulation* (8) is adequate and is widely used in embodied approaches to cognitive science (44). There is strong evidence for mental simulation in mammals (9) and for at least dreaming in birds (10, 94). The theme of simulation will be revisited in the Vertebrates section and it will be suggested as plausible substrate for SE.

My working hypothesis is that there is an (unknown) general function that maps from bodily, including neural, activity to SE. I have evaluated the consistency of proposed theories of SE and the mind with findings in anatomy, experiment, computation, and experience and failed to find any currently proposed theory that is adequate (11). This paper explores whether an evolutionary approach can shed some light on the mystery.

Evolution, especially of humans, is another scientific question of profound importance, but it is not considered to be mysterious. Rapid advances in theory and in a wide range of experimental techniques have led to greatly improved understanding of many of the key evolutionary events and developments (12). There are still unresolved issues and controversies, but many deep questions have been clarified over the last few decades. Perhaps the most famous scientific quote about evolution is Dobzhansky (42): "Nothing in biology makes sense except in the light of evolution."

This is justifiably profound, but is often over-interpreted to imply that biology is totally determined by genetic evolution. Evolution, like gravity, is fundamental property of our world. However, living beings thrive by overcoming and exploiting gravity, evolutionary processes, and other aspects of Nature. One obvious case is niche construction, where animals restructure their environment. Since natural selection acts on phenotypes in their physical and social environment an expanded view of evolution is needed. This formulation, involving epigenetic effects, including culture, is called the Eco-Evo-Devo approach (43). This is the sense of evolution that we adopt here.

By contrast, the mysteries of the mind are becoming increasingly profound as the sciences of brain and behavior advance. Until fairly recently, many scientists, including me, assumed that we would eventually find brain circuits and functions that explain subjective experience. However, as a result of massive cooperative efforts (13), all large areas of the brain already have known functions (Figure 1) so the hope for unknown mechanisms depends on new science.

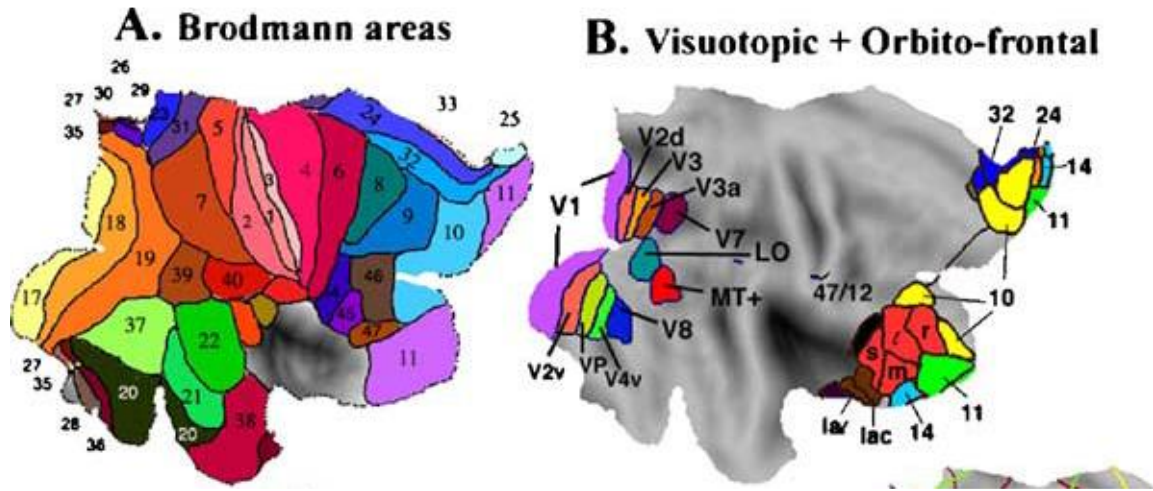

**Figure 1. Flat maps of the Human Brain (13)**

Figure 1A is a standard flattened projection of one hemisphere of the human brain with the various areas colored. The numbers refer to the traditional Brodmann classification of brain regions from their anatomical details. Modern methods (37) have further refined this picture and elaborated the basic functions computed in these different areas. Figure 1B provides more detail on this functional separation in the visual system, which is at the core of the neural binding problem, one of the core mysteries.
.
Figure 1A shows that the functionality of the cerebral cortex is basically known (37) – there is no large available space for current neural computation of currently mysterious phenomena. In addition, various aspects of our visual experience are primarily computed in distinct and often distant interacting circuits. For example, in Figure 1B color calculation is based in the bright green area V4v and motion calculation involves several areas: V3, V3A, MT, etc. In spite of this extreme separation of function, we experience the world as an integrated image with objects that combine all visual properties and even associate these with other senses like sound when appropriate.

The mystery of how all this happens is called the "hard binding problem" (11). This is one of the classic hard problems that continues to motivate current research and theory in SE. No one has proposed a plausible explanation of this mystery; we will reconsider the mysteries of experience in the Conclusions section.

The goal of this article is to approach SE from an evolutionary perspective – how could SE and the mind have evolved? Many distinguished clinicians, experimentalists, theoreticians, and philosophers have worked on ambitious projects on the evolution of mind (14-26). None of these claim complete success in explaining the mind/body problem and neither will I. However, there has been considerable progress on clarifying the issues and providing grounding for continued scientific research.

The first and most fundamental evolutionary criterion is *fitness*. There are many contributing factors to evolution, including considerable randomness. Still, a trait of any living creature tends to survive and evolve according to that trait's contribution to adaptive fitness. For our purposes, the evolution of SE and its precursors offers important constraints on the present function and realization of embodied experience. A crucial concept here is *actionability* (8). Nature determines the fitness of actions, but an organism can only compute actionability, its *internal estimate* of the expected fitness of possible actions, using its internal model (17). There is a well-known theorem of neural computation (97) that shows that any good regulator of a system must contain a model of the regulated aspects of that system. In simple cases like the amoeba or human reflexes, actionability reduces to initiating immediate action. For complex creatures like us, actionability can entail significant exploration and planning and often evokes SE.

Actionability is an extension of Gibson's idea of perceptual *affordances,* https://en.wikipedia.org/wiki/Affordance that adds quantitative, probabilistic, active, situational, goal-directed perception and reasoning. It is also an extension of the popular Bayesian Brain movement (27) to incorporate the requisite *expected utility* (28) of perceptions and actions. Evolution selects on immediate, life-long, and social goals as they contribute to reproductive success, so actionability should include these as well.

The precursors of SE, such as valence (16, 24), which is an organism's evaluation of positive and negative stimuli and perceptual constancies like size and color (19), contribute to actionability estimates and thus directly to fitness. The next Section, Origins, explores some basic life needs of all organisms and how primitive solutions to these needs are often mistaken for core properties of human SE. Almost all of the myriad evolutionary SE theories start with a presumption that the mind will be fully explained by understanding the brain – this is sometimes-called "temporary mysterianism". This position is justified by assertions that the only alternative is to postulate some non-physical effects beyond science. It is true that there is no evidence for SE in the absence of related bodily activity and the possibility remains that other currently unknown physical processes can explain SE. There is my alternative formulation, *agnostic mysterianism* (30) that explicitly acknowledges current mysteries (of the mind, quantum theory, etc.) and also that they may or may not be resolved by current or future science.

This allowance for mystery becomes more important because recent results (11) show that some subjective aspects of everyday experience are **inconsistent** with existing and proposed theories of neural computation. Some of these mysterious experiences are instance are instances of classical mind/body problems like the binding problem (19) and the "illusion" of a detailed stable visual world (11). The inconsistency proofs involve straightforward analysis of common phenomena against proposed neural substrates.

For example, Figure1 shows that visual features are computed in separated brain areas, but we perceive unified objects in full bound detail. Perhaps the most striking example is the SE of continuous motion from sequences of still images in films, videos, and normal vision. This situation is nicely summarized in the following 3/14/19 personal communication from Ken Nakayama:

> *"Despite the fact that this class of Gestalt psychology type demonstrations have been known for over 100 years, the actual physiological basis of any of them is unknown. In fact, it is something that I covered in my large and lengthy review (29) of image motion processing in 1985. I wager that there is no known satisfying physiological understanding of perceived motion experience most generally. This despite the fact that there are hundreds of papers on the physiology and psychophysics of motion."*

In the history of science, such inconsistencies often lead to major conceptual advances. A recent article (30) suggests focusing mind/body research on the border between known and mysterious phenomena and provides some promising examples. Some of these will be discussed in later sections.

The current paper extends the agnostic program of exploring the mind-body problem without presupposing a solution (the brain). In particular, my approach assumes that there is no known solution that is consistent within current neuroscience and that some fundamental breakthrough would be needed. It is also open to the possibility that there are indeed mysteries beyond human comprehension. If this upsets you, consider whether chimpanzees could understand science (31).

As far as I know, none of the many evolutionary approaches to the mind-body problem includes any consideration of the inconsistency with SE of current theories of neuroscience (11). To be fair, all of them stipulate that there are remaining mysteries and some of the articles do contain important insights.

One important common strategy is to separate the mind-body problem, SE, etc. into two distinct sub-problems. The embodied first problem of emotion, perception, etc. is assumed to be tractable and approachable within existing science. The second sub-problem directly involves the mind and the mystery of SE. For example, studies of emotion usually reserve "emotion" for the embodied cases and use "feelings" for the mental aspects (32).

My agnostic approach treats mysteries as the core problem and focuses on computational considerations – how could we have subjective experiences that are inconsistent with current computational neuroscience and how could this phenomenon have evolved? This entails hypothesizing an explicit, unknown function called **X** or Chi that maps from an embodied realization of a phenomenon to its awareness in the mind. The next section, Origins, discusses how some physical aspects of SE occur very early in evolution. The mapping **X** remains a mystery, but evolution suggests some precursors and relevant experiments.

Much of the work referenced here does explicitly employ the term consciousness and I will follow their usage with the understanding that subjective experience is the issue of current concern. Several more or less equivalent terms are employed to describe the *subjective experience* (SE) associated with the mind, but SE appears to be the most focused and least ambiguous. For the core vocabulary, we will consider the following as sharing meaning: *Phenomenology, Subjectivity, 1st person Experience*. Single terms like *awareness, experience,* and *sensation* are sometimes employed, but they have numerous other uses and will not appear here.

The term *Qualia* is also employed, but it has too many versions for our purposes. Since all of these terms (like mind) refer by default to the human mind, we will need another term to denote mind-like behavior in non-human animals. For many purposes, the term *agent* captures the general idea of a living animal. There is more discussion of the importance of language issues in the science of SE in (30). Almost all general discussions of SE focus on either emotions or perception and occasionally both. We will start with these two phenomena, but will examine several other instances of SE, including dreams, extended motor control like a blind man's cane, and anomalous body experiences.

No one seriously argues against the idea that (many) mammals have experiences (14,20,25) that share some of the properties of human perception and emotions like pain, fear, anger, joy, disgust, etc. It is also clear that there are important differences, at least in that only people can talk about such experiences. If we assume that subjective experience has evolved then there must be evolutionary *precursors* that also need to be studied (16, 20, 26). Evolutionary considerations are the core concern of this paper.

One useful distinction is the recent C0, C1, and C2 definitions in Dehaene et al. (6). They define C0 as unconscious neural activity that is not experienced, C1 as activity that is accessible for computation and report, and C2 as self-monitoring of C1.
The discussion in (6) focuses on some simple information processing functions of C1 and C2, but the classification is much more general. One benefit of this suggestion is that the symbols C0,1,2 do not come with any conceptual baggage. Another advantage of this proposal is that it provides a label (C0) for the vast range of behavior in the absence of awareness. The functions of C0, C1, and C2 are not mutually exclusive and often overlap, for example, when you stumble while walking.

Human C0 (6) behavior ranges from reflexes and other intrinsic circuits to homeostasis and language understanding. In fact, much human activity (e.g. driving) requires automatization to C0 and cannot be done mindfully as C1. In this terminology, SE corresponds roughly to C1 and each involves both the embodied *emotions* and the mental *feelings (32)*. A complication is that automatized behavior (C0) can give rise to SE. For example, unconscious interoception systems evoke SE of hunger, fatigue, etc. Emotions are normally automatic, but definitely evoke SE (32, p. 83))

There is general agreement that subjective experience (SE) is an essential aspect of the mind and of any notion of consciousness - we will focus on SE unless specifically stated otherwise. There is considerable evidence that some animals have C1 but not C2. C2 does seem necessary for language - for example to express "I used to like candy". The many important functions of C2 are beyond the scope of this paper.

However, a huge problem is that SE is essential but still mysterious. Researchers in fields such as emotion and perception continue to make progress, often by deferring any questions involving subjective experience. A significant fraction of the research on emotions and on perception explicitly excludes SE, acknowledging the present mysteries. A striking recent example is the magisterial book by Adolphs and Anderson entitled "The Neuroscience of Emotion" (32). The authors choose this title because they explicitly omit any considerations of SE. As part of their justification for this, they invoke the great success of this separation strategy in perception research. The Nakayama quote above on motion perception makes the same point about the remaining mystery of the bodily realization of perception.

More generally, it is essential to distinguish the two distinct general aspects of SE. Any SE in perception, emotion, motor control, etc. has an embodied substrate that may or may not be understood. Even when the embodied basis of perception and emotion is understood, many of the ancient mysteries have no known embodied realization (11). As proposed above, this entails a (mysterious) function $X$ or Chi (30) that maps from bodily activity to subjective experience and is the core of the Mind-Body problem.

Our discussion in the next section will include protozoa with no neurons, so "neuroscience" is not always the appropriate term. The general form of the mystery is simply: "$X$ maps from bodily activity to related SE". To avoid unnecessary pedantry, we stipulate that neural firing is the dominant component of $X$ bodily activity in vertebrates. All of our discussions of particular SE will explicitly distinguish between the (possibly understood) bodily substrate, including hormones and the microbiome, of phenomena and the subjective mental component involving $X$. Any question of a causal mapping from the mind to the body is ruled out by the massive influence of bodily activity, including homeostasis, on the mind. Top down mental control would also involve theological issues beyond the scope of (current) science. It is unscientific to postulate causal effects of a "mind" that has no substance.

I also use the terms *evolution* and *fitness* in the conventional manner as they apply to all living things. We are adopting the broad notion of evolution that includes niche construction, cultural change and development that is called the Extended Evolutionary Synthesis. https://en.wikipedia.org/wiki/Extended_evolutionary_synthesis. This includes consideration of group, niche, social, and epigenetic effects in evolution.

There is no way for an organism or a scientist to calculate present fitness, since it depends on the future. We introduced the term *actionability* above to label an organism's probabilistic internal estimate of the expected fitness of its potential actions in the current situation. Any living system will include such an internal mechanism for choosing actions. This will incorporate an estimate of *valence* – whether something is good or bad for the organism.

Damasio (18) and LeDoux (24) suggest that primitive valence in simple creatures like bacteria and amoeba is the evolutionary foundation of all emotion.

It may seem too ambitious to study the evolution of subjective experience in the absence of a scientific explanation of the phenomenon of SE itself. Paul Cisek (33) makes a strong case for starting from an evolutionary standpoint in what he calls "phylogenetic refinement" and illustrates this approach with a detailed history of feedback architecture in the development of the brain. Joseph LeDoux (24) presents a related plausible evolution of subjective experience, emphasizing protection circuits as the foundation of emotion. His Chapter 27 includes a fascinating story on how neurons and nervous systems could have evolved, in multiple steps, from the sensing and acting chemical reactions in protozoa.

Any naturalist account of SE entails an evolutionary story and discussion of possible precursors, including protozoa. There is a crucial methological issue for SE that always arises – perspective. Most science is inherently public and is based on a general (3rd person) perspective, accessible to anyone who has adequate knowledge, and enforced by social practices.

However, several related intellectual developments (34, 35) have argued effectively that the subjective (1st person) perspective is crucial for any science of SE and the mind. The "1st person" terminology is, by definition, inappropriate for non-human animals, robots, etc. A generally useful alternative concept is "agent perspective". We will also need to extend the standard definition of "percept" – *"the mental result or product of perceiving, as distinguished from the act of perceiving; an impression or sensation of something perceived".* For us, a (generalized) *percept* is an embodied product of processing sensory inputs that increases evolutionary fitness. As we will see, such computed percepts occur early and often in evolution. In at least humans and some other vertebrates, percepts can cause SE.

Independent of any particulars, the crucial fact about SE is the focus on experience from the perspective of an experiencing creature. This agent perspective is also popular as Enactivism, a component of the current 4E embodied, embedded, enactive, and extended cognition movement [https://4ecognitiongroup.wordpress.com/](https://4ecognitiongroup.wordpress.com/) . However, the 4E movement (36) is too general for our study of the evolutionary precursors of SE. In the next section, we will explore specific proposals including autopoesis, corollary discharge, actionability, and the umwelt approach, as a basis for studying agent perspective in simple animals.

The philosopher and Octopus fan, Peter Godfrey-Smith (PGS), has written several papers on evolution and the mind-body problem. A recent one of these (23) shares many of the attitudes and ideas with this paper, but also has one major difference.

Both efforts recognize the current mystery of SE and do not claim that studying its evolution will produce a major advance. They also both ground the project on the twin anchors of experienced subjectivity and the simplest animals. PGS (23) has a succinct but strong discussion of the philosophical background for Levine's "explanatory gap" between SE and any scientific substrate. His second section introduces the notion of "subjectivity" focuses on its importance for the conceptual analysis of SE. The key idea is that analysis must be from the perspective of a subject, which can be thought of as an animal. He then uses this formulation to consider several current philosophical arguments about the mind-body problem.

Exploiting his position as a philosopher, PGS constrains the scope of his effort:
"Levine also presented his gap through a demand for an explanation of why the feel of seeing *red*, for example, goes with *this* brain state. I don't offer explanations of this kind – those explanations are a task for neurobiology. "

This where we part company - for me the relation between SE and causal brain/body states (embodiment) **is** the mind-body problem. Nevertheless, PGS' third section, on Evolution, discusses many relevant facts and issues and will be cited below. PGS works hard to be neutral, but he starts and ends as a committed materialist/physicalist. He does allow that the notion of "physical" is likely to change significantly in the future so his hope is that the mind-body problem will be resolved by something that can be called physical. This is indeed a consummation devoutly to be wished, but there is no evidence of it.

A radically different view of evolution and the mind appears in "The Case Against Reality" by Donald Hoffman (41). He proposes a "conscious realism" in which only the mental is real and where evolution has duped us into our beliefs about material reality. The book's key technical contribution is the Fitness Beats Truth (FBT) Theorem. This states that evolution essentially always favors fitness over truth. As Hoffman says, the theorem is quite abstract and follows from the formal definition of Universal Darwinism:
https://en.wikipedia.org/wiki/Universal_Darwinism. The result does not depend on any particular instantiation. The catch in the FBT theorem is that it provides no guidance on how to *find* the fitness features. The basic result is circular – Universal Evolution favors whatever traits improve fitness. This consistency result could be of heuristic and philosophical value, but does not help the scientist or the organism.

*A more recent contribution (88) is* "*The evolution and development of the uniquely human capacity for emotional awareness: A synthesis of comparative anatomical, cognitive, neurocomputational, and evolutionary psychological perspectives". This discusses some of the same issues as the current article, from a different perspective, as they state: "Emotional Awareness EA, as operationalized in this paper, is primarily measured linguistically."*

*This linguistic focus on EA is narrower than our SE and obviously entails concentrating on precursors to human abilities: "The level of EA displayed by non-human animals appears limited to experiential qualities such as valence, arousal, and motivated action". The paper includes extensive discussions of what they view as the crucial precursors of EA, including brain size and structure, co-opting of early biochemical functions, and the hyper-social nature of human groups. All of this is also important for SE. They present significant analysis of simulation and of computational models, based on contemporary Bayesian free-energy theory (89). There is also examination of alternative formulations and open questions. Another recent overview is (92)* Subjective Experience and Its Neural Basis. *This book chapter surveys a wide range of issues related to SE, but is admittedly early days.* "At present, however, the emerging picture sketched in this chapter already serves to highlight a rich tapestry of inter-related states and processes underlying our internal experience. "

One ancient and continuing mystery of subjective experience is why it feels so different from anything else that we know about the world, including our own body. A related, and much more tractable, problem is the evolutionary fitness advantage of subjective experience. This is discussed in the next section, Origins, starting from related adaptations in primitive creatures.

One of the clearest examples of evolutionary adaptation is what is called "perceptual constancy" or "subjective constancy", particularly in vision. Our perceptual systems compute values of several features that are not actually constant, but they do approximate the external (distal) world based on signal (proximal) values **https://en.wikipedia.org/wiki/Subjective_constancy**.

For example, we compute fairly accurate estimates of the size, distance, and motion of external physical objects and this is obviously crucial in acting appropriately. Quite a lot is known about the neural circuitry and computation underlying these constancy calculations in many animals (38). What remains mysterious is our subjective experience of a fully detailed and physically accurate scenario. There is no known neural substrate for this subjective phenomenon and considerable evidence (11) that these SE are inconsistent with any known or proposed theory of neural computation.

Another fundamental visual constancy involves color. Many animals can act upon the basic reflectance of objects in the world largely independent of changes in the ambient light, for example between broad daylight and dusk (39). Again, this has obvious fitness advantages for many animals, including in insects, as we will see. So perhaps the basic selectional advantages of calculated perceptual features (percepts) could shed light on precursors of subjective experience – this will be discussed in the next section.

**2. Origins of Subjective Experience**

Evolution encompasses all life on earth, but here we are especially interested in precursors to human subjective experience (SE). It seems appropriate to start with animals and their prototypical protozoan ancestors the Amoeba. There is a wide range of amoebic forms including our white blood cells. Free living "Amoeba Proteus" are the most commonly studied amoeba and are prototypical single cell creatures. They are described by Damasio (15).

"All living organisms from the humble amoeba to the human are born with devices designed to solve automatically, no proper reasoning required, the basic problems of life. Those problems are finding sources of energy; incorporating and transforming energy; maintaining a chemical balance of the interior compatible with the life process; maintaining the organism's structure by repairing its wear and tear and fending off external agents of disease and physical injury. "

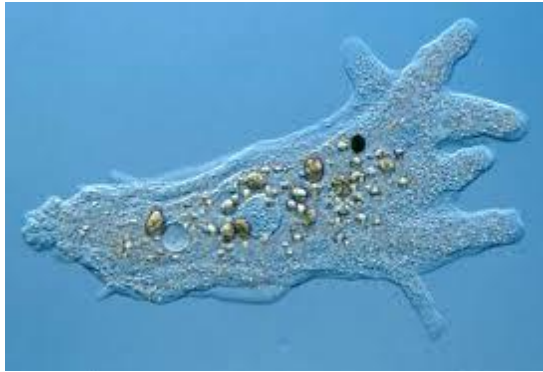

Figure 2. Photomicrograph of an Amoeba

The Amoeba, is described in: https://en.wikipedia.org/wiki/Amoeba_proteus

"*This small [protozoan](#) uses [tentacular](#) protuberances called [pseudopodia](#) to move and [phagocytose](#) smaller [unicellular](#) organisms,(which may be greater in size than an amoeba), which are enveloped inside the cell's [cytoplasm](#) in a food [vacuole](#), where they are slowly broken down by enzymes. Amoeba proteus is very well known for its extending pseudopodia. It occupies freshwater environments and feeds on other protozoans, algae, rotifers, and even other smaller amoebae.*"

In his recent book (18), Damasio discusses even simpler creatures, bacteria, and their remarkable individual and group behaviors. The crucial point for Damasio and for this article is the *continuity* of the requirements of life and the evolution of fitness. Damasio frames this process as an extension of homeostasis:

"*Bacteria can sense the conditions of their environment and react in ways advantageous to the continuation of their lives. … much later in evolution neurons and neural circuits have come to make good use of older inventions that relied on molecular reactions and on components of the cell's body* "(18, p.54).

Joseph LeDoux, who focuses on the evolution of emotions and feelings, stresses the importance and ubiquity of *valence* for all living things. All creatures need to incorporate notions of helpful and harmful conditions and respond appropriately (24).

"The survival circuit concept integrates ideas about emotion, motivation, reinforcement, and arousal in the effort to understand how organisms survive and thrive by detecting and responding to challenges and opportunities in daily life."

*Perspective* is another crucial issue for SE that arises even in these discussions of bacteria, amoeba, and other simple creatures. The choice of perspective can have a profound influence on how phenomena are considered. Most science is inherently public and is based on a general (3${}^{rd}$ person) perspective, accessible to anyone who has adequate knowledge, and enforced by social practices. However, several related intellectual developments discussed below have argued effectively that the subjective (1${}^{st}$ person) perspective is also crucial for any science of SE and the mind.

The 1st *person* terminology does not make sense for non-human animals, robots, etc. A generally useful term here is "agent (internal) perspective". Independent of any particulars, the crucial fact about SE is the focus on experience from the perspective of an experiencing creature. Both the Damasio and Wikipedia discussions of Amoeba were from this agent perspective and the LeDoux survival circuits above are clearly agent oriented. Efforts to build robots or other autonomous systems are inherently agent focused.

Two early and complementary scientific manifestos jointly provide an intellectual foundation for a science of agent experience from protozoa to human subjective experience. The dependence of a creature's life on its surroundings is stressed in the "umwelt" literature (https://en.wikipedia.org/wiki/Umwelt). The German term "Umwelt" can be partially translated as "surroundings". More technical is the Oxford dictionary – "the world as it is experienced by a particular organism". The Umwelt will vary from agent to agent, as in this original von Uexküll example:

 "Every object becomes something completely different on entering a different Umwelt. A flower stem that in our Umwelt is a support for the flower becomes a pipe full of liquid for the meadow spittlebug (Philaenus spumarius) who sucks out the liquid to build its foamy nest "

One important consequence is that animals can only partially perceive their environment. In addition, many aspects of the physical and social environment are non-deterministic so fitness must address this reality. Probabilistic internal models play a major role in human SE (17) and will be discussed in Section 5, People.

The key idea is that all living things must incorporate mechanisms that relate what they can try to sense to what they can try to do. This is foundational in all animals, including people and we will use general term *Internal Model* (17) for this function; this will be discussed in detail in Section 5. One crucial insight is that the internal model required for an animal's fitness is not a veridical representation of the world (41). One general form of the mind-body problem is: How does an animal's actions and its internal model affect its perceptions?

Sensory input can be separated into two streams: afferent information, which is information that comes from the external world, and reafferent information, which is sensory input that arises from our own actions. One proposed advantage for internal models is that they help us to determine whether a sensory input we receive is a consequence of our own actions. They also serve to filter out the components of input arising from our own actions. Even an amoeba reacts differently to external physical contact resulting from its own actions.

In higher animals, this adaptive mechanism is called "corollary discharge" CD (45) , and is exemplified in the sonar of bats. A bat needs to know what sonar signals it has emitted in order to make sense of reflected inputs. Corollary discharge (aka efference copy) is also involved in the stable image over saccades that was our initial example. If you gently move your eyeball, the scene will shift and not remain stable. CD helps explain the neural basis for the "illusion" of a stable visual world, but the SE itself remains mysterious.

The Umwelt story itself does not directly address the internal realization of behavior or mechanisms involved. The internal organization of all living things is exactly the focus of the *Autopoesis* literature - https://en.wikipedia.org/wiki/Autopoiesis. The term was coined by Maturana and Varela and is translated as "self-creating".

"*Autopoiesis was originally presented as a system description that was said to define and explain the nature of living systems. A canonical example of an autopoietic system is the biological cell. The eukaryotic cell, for example, is made of various biochemical components such as nucleic acids and proteins, and is organized into bounded structures such as the cell nucleus, various organelles, a cell membrane and cytoskeleton. These structures, based on an external flow of molecules and energy, produce the components, which, in turn, continue to maintain the organized bounded structure that gives rise to these components. (Not unlike a wave propagating through a medium).*

*Moreover, an autopoietic system is autonomous and operationally closed, in the sense that there are sufficient processes within it to maintain the whole. Autopoietic systems are "structurally coupled" with their medium, embedded in a dynamic of changes that can be recalled as sensory-motor coupling This continuous dynamic is considered as* a rudimentary form of knowledge or cognition and can be observed throughout life-forms."

The crucial point for us is that the autopoesis and the umwelt insights together provide a coherent agent-focused foundation for the internal and external constraints of life. The details obviously differ widely across animal experience, but the umwelt principles remain.

We will focus on a broad, but bounded range of animals. The lower bound will be protozoa, prototypically amoeba. For the upper bound, we will assume that full "subjective experience" requires a "subject" or "self" (15, p10; 22) and that the only known subjects are human, prototypically a scientist with some interest in the Mind-Body problem. Tomasello (31) has explored and described profound differences in the mind and behavior of humans and those of our nearest primate relatives. This will be discussed in Section 4, Primates. Even protozoa need to be considered as agents and thus proto-subjects. That is, amoeba are living beings that react individually and socially to internal state and some external activity that they sense.

A major challenge in this effort is to disentangle parallel behaviors from actual evolutionary causality. The core idea is that all living creatures have many shared fitness requirements and will thus have analogous structures and behaviors. Similar features found in disparate animals could be due to convergent evolution https://en.wikipedia.org/wiki/Convergent_evolution or a most recent common ancestor (MRCA) https://en.wikipedia.org/wiki/Most_recent_common_ancestor, or both. Convergent evolution results from similar fitness requirements. A striking example is the extensive homologies in shape and function between aquatic mammals and fish. For another example, vision is believed to have evolved several separate times. My goal here is to explore what is known about precursors of human subjective experience SE and C1 (awareness), as discussed in the Prologue.

Specialized abilities can serve essentially the same function as much more general capabilities in "more advanced" organisms. It is essential not to assume that certain complex behavior requires particular embodiment, including subject-hood. A particular problem that I will discuss is a current fashion of using terms like SE and consciousness to describe the behavior of a wide range of animals, including insects and cephalopods. This is a standard concern and goes by the name of Morgan's (1894) canon:

*"In no case is an animal activity to be interpreted in terms of higher psychological processes if it can be fairly interpreted in terms of processes which stand lower in the scale of psychological evolution and development."*

In Morgan's time, much less was understood about brain and behavior and the canon was just about *interpretation* of behavior because virtually nothing was known about the details of neural structure and resulting behavior. A current canon would caution against using human features to describe behaviors that can be explained more directly in terms of the animal itself. The core problem in the evolution of SE is distinguishing primitive functional capabilities from mammalian level SE.

One striking property of SE is the *inconsistency* of subjective perception with the known physiology, computation, and behavior of the nervous system (11). Focusing on vision, we are familiar with many visual "illusions" and mysteries like those described in the Prologue. As discussed there, the perceptual "constancies" are among the most familiar and ecologically important instances of human subjective experience. We also introduced there our use of the term "percept" to mean the result of some internal processing of sense data, including internal sensing and corollary discharge. The (visual) percepts needed for effective action are not a straightforward function of the sensory input and much of visual science studies this.

The requirement for additional computation between sensory input and actionable percepts is also very widespread among animals. Most basically, an animal must consider its own actions when evaluating sensory inputs. The internal signal of action is still sometimes called "efference copy" but is usually denoted more generally as "corollary discharge" or CD (45).

*"The CD taxonomy consists of higher- and lower-order categories that are based on the operational impact of the signal on the nervous system. Lower-order CD signaling is used for functions such as reflex inhibition and sensory filtration, whereas higher-order signaling participates in functions such as sensory analysis and stability, as well as sensorimotor planning and learning. "*

Inhibition mediated by CD enables reflex coordination in diverse animals such as nematodes, tadpoles and gastropods.

*" CD for sensory analysis and stability enables organisms such as the macaque, the rat, and the bat to move and yet experience the world as it is (stable and continuous) rather than as it is sensed at the receptor level (in a chaotic and piecemeal fashion)."*

CD is ubiquitous and a crucial part of SE, but it is not the hard part. The fitness advantage is obvious, but there is still no story about why (at least we humans) have the SE of the binding problem, etc. (11). Therefore, the core mystery remains Chi, the postulated mapping from bodily activity to SE. In the next section, we will consider what is known about C1 (awareness) and SE in vertebrates.

We can further illustrate the situation using our running example of color constancy. It should be no surprise that other animals benefit from an ability to determine the reflectance of useful objects under varying illumination conditions. Many animals have effective color constancy. Moths are not usually thought of as especially intelligent (cf. bees) but their use of color and other cues has been studied extensively (39).

The moth's computed color is an example of an evolutionary early *percept* as discussed in the Prologue. Perceptual color is the basis for a moth finding the appropriate flowers. Hawkmoths possess color constancy: *M. stellatarum* and *D. elpenor* recognize the same flower color under changing illumination spectra. This ability is especially important for foragers that are active under a range of lighting conditions, such as sunlight and shade, and during dawn and dusk.

Receptor adaptation contributes a large part to constancy. The sensitivity of the receptors decreases as a result of adaptation of the photoreceptor cells being stimulated by the background illumination. For chromatic adaptation, it is assumed that the different receptor types adapt separately depending on the background spectrum. Chromatic adaptation can be described by the von Kries coefficient law (https://en.wikipedia.org/wiki/Von_Kries_coefficient_law**)**, which scales the signals from the photoreceptors to the background illumination to keep the perceived color constant despite changing illumination spectra.

Hawkmoths have trichromatic color vision based on three spectral receptor types sensitive to ultraviolet, blue, and green light. Quantitatively, the sensitivity is between 10 and 100 times higher for blue light than for lights of long wavelengths. This cannot be understood from receptor properties including receptor noise alone, but indicates top-down neural regulation processes that control color salience.

Adaptation is another moth mechanism that supports constancy over intensity changes. In the dark-adapted state, a clear zone allows light from a given direction to pass through any of several hundred facets to reach the photoreceptors in a single sensor. This greatly improves visual sensitivity in dim light. In bright light, screening pigment migrates into this clear zone turning the superposition eyes into functional apposition eyes and adapting the visual system over several orders of magnitude of light intensity. The moths can also adapt to pursue colors (e.g. green) that are not innately favored or present in their natural habitat (39).

This remarkable color ability of Moths is sometimes used in arguments that (some) insects have abilities that should be labeled as a form of consciousness (46, 47). Some octopuses also exhibit remarkable abilities (22), but Morgan's canon suggests caution in attributing these to mammal-like SE or consciousness. As we discussed above, Morgan's canon and its updated version caution against this.

More generally, in a standard phylogenetic tree (Figure 3) we can see a very early divergence of vertebrates (Chordata) from insects (Arthropoda) and octopuses (22) (Mollusca). The remarkable capabilities of bees, octopuses, etc. including learning and memory , evolved concurrently well after the evolutionary split from the branch leading to vertebrates (Chordata) and us and so are unlikely to be directly informative on human SE.

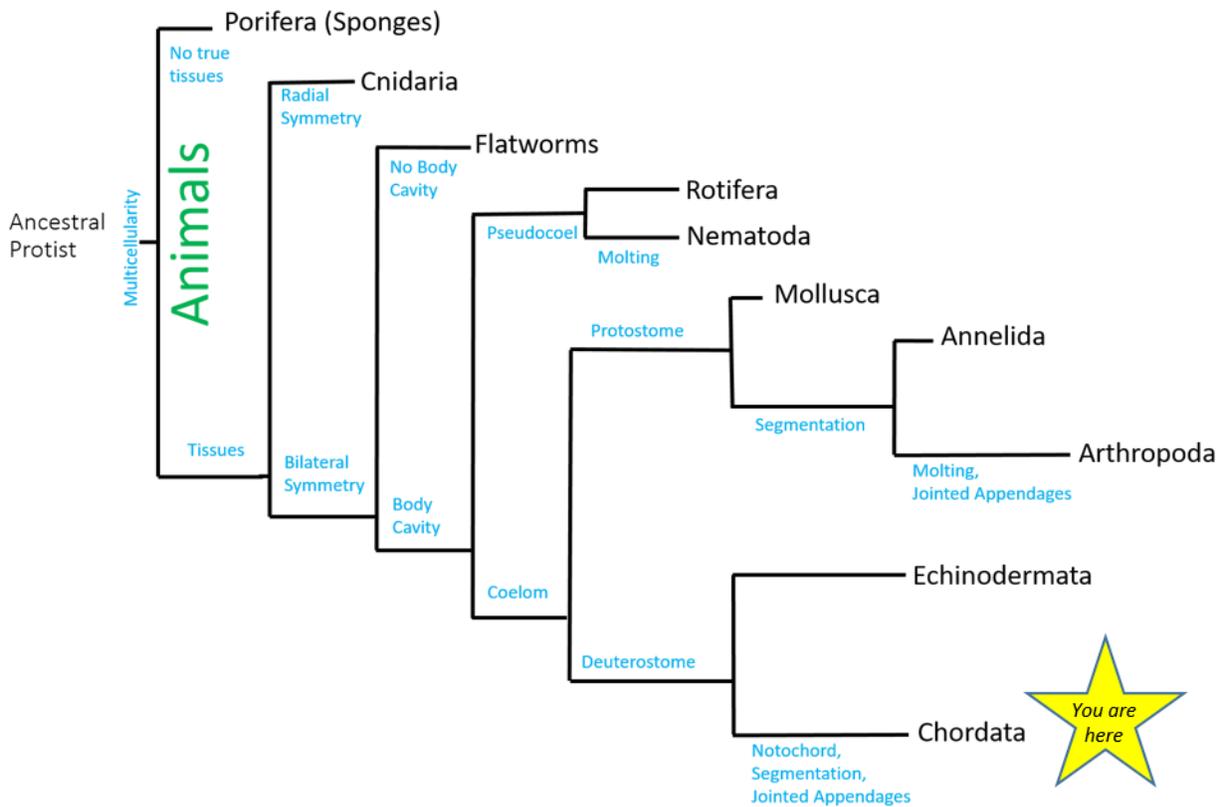

Figure 3. Animal Clade Tree

We will continue exploring the evolution of subjective experience in the following Section 3, Vertebrates. The main line goes through Chordata and seems to require some primitive brainstem mechanisms that are preserved even to humans (26).

## 3. Vertebrates and Simulation

We have seen in the previous Origins section that even the simplest animals exploit internal models and processing to convert sensory and other inputs to adaptive action.

This is part of what is entailed by human subjective experience, SE, but not the whole story. The additional factor that makes the experience *subjective* goes by many other names including qualia, awareness, phenomenology, etc. and is a foundation of any notion of consciousness, as discussed in the Prologue.

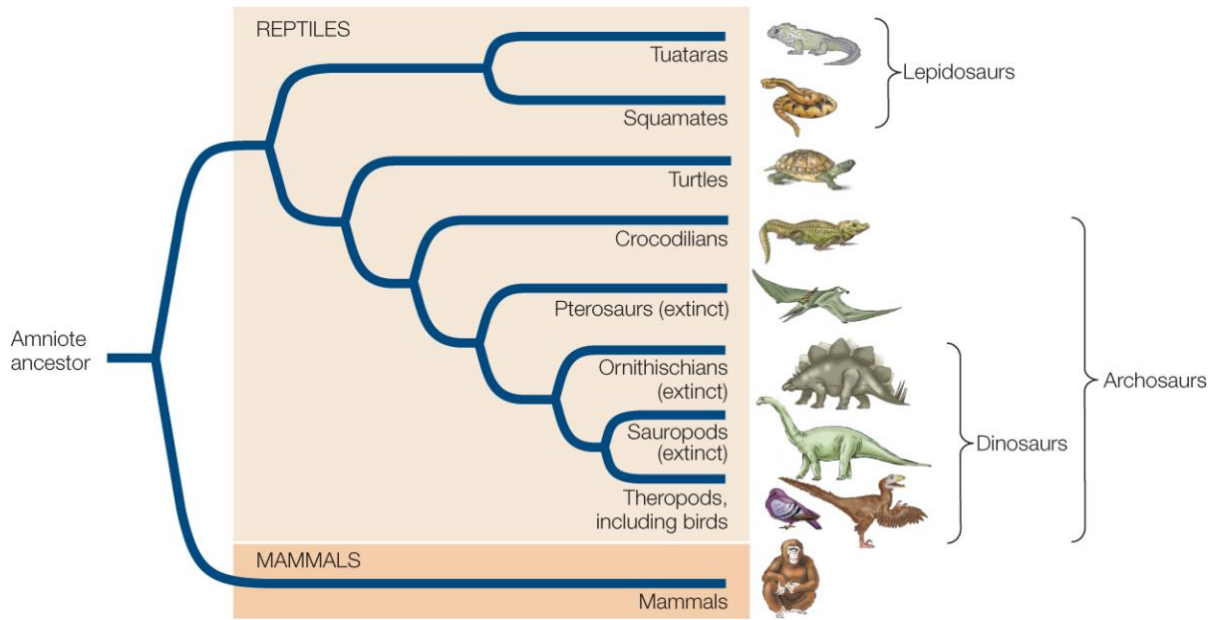

Figure 4 Vertebrate Evolutionary Tree

Although it is far from consensus, the dominant scientific opinion places the direct system precursors of human SE in *vertebrate* evolutionary history (Figure 4)**.** In this section, we will provide evidence that a form of SE evolved in parallel in mammals and birds. Recent findings, including replay in dreaming, suggest that mental *simulation* (8) is a plausible basis for SE. There are certainly much older chemical and genetic factors in play, as discussed in the Origins section. That previous section outlined some impressive invertebrate abilities related to SE and introduced Morgan's canon (62) that warns us against assuming that such behaviors should be taken as evidence for vertebrate mechanisms. We also saw that the remarkable capabilities of bees, octopuses, etc. evolved concurrently well after the evolutionary split from the branch leading to vertebrates and us (Figure 2) and so are unlikely to be directly informative on human SE.

 A stable distinctive trait, like SE, generally evolves through many steps, only some of which are understood. The Last Common Ancestor (LCA) of birds and mammals is known to be *amniote*s (Figure 4), the first vertebrates to reproduce on land. Immediate precursors include the ragworm, which had the first tripartite brain, but lacked a spinal cord and the lamprey which is considered the first vertebrate. The rudimentary brainstem's motor programs remained confined in these creatures to executing protective reflexes as well as eye and body movements (12). Several additional evolutionary steps have been discovered.

Hagfish possess a simple thalamus which integrates sensory information. The evolution of such a structure that could combine sensory information from multiple sources would have been extremely beneficial, providing a multimodal sensory experience of the world. As other brain structures evolved, the brainstem became critical for generating general states of systemic arousal, reward, and stress (12).

Much of the core brain stem structure and chemistry of these amniotes has been preserved even in people, albeit with considerable evolutionary change. Therefore, one baseline for the possible evolution of human SE is the amniotes. An important feature of amniote brains is the centralization of perception and action in the brain, replacing many individual reflexes with a centralized valence system. Among other things, this brain centralization supports much richer internal models and functions, including simulation (26). Model based simulation is often suggested as a precursor to human SE and Consciousness and this will be a focus of this section.

The upper bound for this section on vertebrates is the broad range of SE capabilities exhibited by both mammals and birds (94). These include convincing evidence for the standard SE examples in perception and emotion, but also dreams, etc. For example, no one seriously claims that dogs lack emotional and perceptual experience. The following section, Primates and Embodiment, considers evidence for additional abilities in primates and suggests some relevant experiments. This article does not directly address exclusively human capabilities, including language and reflective awareness, called C2 in the Prologue (6). However, SE in humans plays a central role in the People section and the Conclusion.

You can see from the vertebrate clade chart (Figure 4) that the last common ancestor (LCA) of birds and mammals are the *amniote*s, characterized by breeding on land. Recent findings https://en.wikipedia.org/wiki/Evolution_of_mammals are starting to fill out various intermediate forms along the mammalian branch. Much of the vertebrate neural structure is carried over to people and seems to be necessary for human SE. In general, similar traits can result from convergent evolution or from a common ancestor (LCA) or both. An excellent summary of the diverging evolutionary paths to birds and mammals is https://en.wikipedia.org/wiki/Synapsid).

For our purposes, it is best to start with Merker's (26) magisterial review *Consciousness without a cerebral cortex: A challenge for neuroscience and medicine.* This BBS article presents an authoritative description of the general vertebrate brain-stem architecture and function. Merker's direct focus is on consciousness, but everything also holds for SE.

"This upper brain stem system retained a key role throughout the evolutionary process by which an expanding forebrain – culminating in the cerebral cortex of mammals – came to serve as a medium for the elaboration of conscious contents. This highly conserved upper brainstem system, which extends from the roof of the midbrain to the basal diencephalon, integrates the massively parallel and distributed information capacity of the cerebral hemispheres into the limited-capacity, sequential mode of operation required for coherent behavior. It maintains special connective relations with cortical territories implicated in attentional and conscious functions, but is not rendered nonfunctional in the absence of cortical input. This helps explain the purposive, goal-directed behavior exhibited by mammals after experimental decortication, as well as the evidence that children born without a cortex are conscious. Taken together these circumstances suggest that brainstem mechanisms are integral to the constitution of the conscious state, and that an adequate account of neural mechanisms of conscious function cannot be confined to the thalamocortical complex alone."

The many accompanying BBS commentaries on this article agree with the basic assessment, and mainly stress that the thalamus and cortex also play a major role in human SE and consciousness. *The function of cortical and sub cortical activity in SE (and Cs.) are well characterized in (98).*

*" Cortical lesions can thus result in such specific impairments of consciousness that one may no longer be able to speak, perceive color, or identify parts of themselves as their own. Damage to lower midline brain structures, on the other hand, will likely alter the level of consciousness (i.e., arousal) without necessarily changing its contents. "*

There is now general agreement that the amniote brain-stem mechanisms are conserved in all vertebrates and play a key role in internal mental states, including SE. Therefore, this is a plausible lower bound on precursors to human SE. We start with some reports of SE-like phenomena in intermediate vertebrate forms like fish and some lizards. Fish constitute the earliest amniotes and, at least arguably, exhibit experiences related to human SE. An obvious focus is *pain* in fish, which remains controversial. There is general agreement that parts of the pain networks have not changed significantly in vertebrate evolution.  https://en.wikipedia.org/wiki/Pain_in_invertebrates

"Although there are numerous definitions of pain, almost all involve two key components. First, nociception is required. This is the ability to detect noxious stimuli, which evokes a reflex response that moves the entire animal, or the affected part of its body, away from the source of the stimulus. The concept of nociception does not imply any adverse, subjective feeling; it is a reflex action. The second component is the experience of 'pain' itself, or suffering—i.e., the internal, emotional interpretation of the nociceptive experience. Pain is therefore a private, emotional experience."

So, here again we have the distinction between some underlying bodily function and the related internal subjective feeling. There is almost no suggestion of emotional pain in invertebrates, and the case for fish is highly contested. Since fish are the most primitive vertebrates and do have the basic vertebrate brain-stem structure, fish are a possibility as a precursor to some human emotions. Independent of any scientific issues, the question of pain in fish is a continuing social concern (48), but this will not be addressed here. *More generally the review (48) is the most complete survey of the shared structure and behavior of all vertebrates as a basis of human SE. This survey does not examine the additional capabilities of mammals and primates that are addressed below.*

In addition to emotions, visual perception has been the major focus of SE research. There is increasing attention to vision in non-human, especially vertebrate animals. Historically, much of what we know about visual circuitry has come from studies of mammals, especially cats and monkeys. An excellent recent overview journal issue is "What can simple brains teach us about how vision works"
https://doi.org/10.3389/fncir.2015.00051

The earliest vertebrates, fish, already show several important features including color constancy as we saw for moths, as well as additional features (49).

"We have reviewed studies that reveal the mechanisms used by the visual system of fish for adaptive object perception. The fundamental functioning principles that allow the appreciation of objects as unified entities, segregated from the background and characterized by invariant properties seem to be shared between species belonging to distant vertebrate classes"

"Teleost fish represent an ideal model to identify basic information processing mechanisms that provide the functional building blocks of social behavior across different species with varying social systems."

In addition, fish respond like people to some common visual illusions (50) :
"we show that fish perceive one of the most studied motion illusions, the Rotating Snakes. Fish responded similarly to real and illusory motion."

Again, this is informative about bodily visual function, but silent on SE. Moving on, many birds, particularly raptors, have outstanding visual abilities, but the studies related to SE have been mainly focused on chickens and pigeons. Of great interest here is (51).

" In the Evolution of Mechanisms of Object Recognition in Vertebrates: A Working Hypothesis, (we) will propose our current working hypothesis regarding the evolution of object recognition mechanisms in vertebrates, aiming toward explaining similarities and differences between pigeons and people (and other primates) found in behavioral studies."

Again, all of these findings concern the behavioral and functional aspects of visual processing, but do not directly address SE. Unsurprisingly, the range of SE manifestations in birds (94) and mammals are much richer than in fish, lizards, etc.

Wu (52) says:
 "For vertebrates, extant birds and mammals share a number of highly similar characteristics, including but not limited to, enhanced hearing, vocal communication, endothermy, insulation, shivering, respiratory turbinates, high basal metabolism, grinding, sustained activity, four chambered heart, high blood pressure and intensive parental care. These bird-mammal shared characteristics (BMSC) are considered to have evolved convergently in the two groups. Given their similar adaptation to nocturnality, we propose that the shared traits in birds and mammals may have partly evolved as a result of the convergent evolution of their shared early ancestors."

One particularly interesting phenomenon is dreams. Mammals, birds, reptiles, amphibians, and some fish exhibit sleeping behavior. Some form of sleep appears in insects and even in simpler animals such as nematodes. In the behavioral sense, sleep is characterized by minimal movement, non-responsiveness to external stimuli, the adoption of a typical posture, and the occupation of a sheltered site, all of which is usually repeated on a 24-hour basis. Insects go through circadian rhythms of activity and passivity but do not seem to have a homeostatic sleep need. Insects and other invertebrates do not exhibit dreaming or REM sleep.

Mammals, birds and reptiles evolved from common amniotic ancestors, the first vertebrates with life cycles independent of water. The fact that birds (94) and mammals are the only known non-human animals to exhibit REM and NREM sleep suggests a common trait before divergence. Reptiles are therefore a logical group to investigate for the origins of sleep and dreaming. Daytime activity in reptiles alternates between basking and short bouts of active behavior, which have significant neurological and physiological similarities to sleep states in mammals. It has been proposed (53) that REM sleep evolved from short bouts of motor activity in reptiles and Slow-Wave Sleep (SWS) evolved from their basking state, which shows similar slow wave EEG patterns.

The general fact that mammals and birds share many properties, amniote LCA, and convergent features suggests that sleep is a good foundation for the study of the evolution of SE. Simulation during sleep is a crucial component of memory consolidation (83). More specifically, vertebrate dreaming in REM sleep seems similar to human dream behavior and could be a precursor to SE. The SE of dreams is produced from bodily activity that is (largely) generated internally and is thus plausibly related to my hypothesized Chi mapping from bodily activity to SE. Dreams are also an instance of simulation and are often described as such. In fact, one prominent theory of dream evolution is based on the simulation of potential threats (54).

**Simulation**

*Simulation* is another term with many meanings, one of which plays a fundamental role in this article and more generally in theories of the mind. The contemporary meaning of simulation developed along with the theory and practice of computing. For people (like me) with a computational background, this involves models of various systems and phenomena as digital computer programs. The "embodied simulation" theory of mind used in this paper (8) proposes that mental agents rely heavily on their own internal (neural) simulations for dealing with the physical and social world.

From a technical computation perspective, simulation adds capabilities unavailable with static models. As we have discussed, all living agents need "internal models" to map from goals and perceptions to actions. Many organisms also have capabilities for learning and memory that enable them to build and exploit new internal models. Simulation adds *detachment*, a powerful novel ability to evaluate variants of past experience and the possible consequences of future actions. This is central to my theory of the evolution of SE and will be revisited in the Conclusion section.

There is a largely independent, but consistent, exploration of mental simulation in the cognitive sciences. In an oft-quoted 1943 passage, Craik ( 55 pp. 59-61):

"If the organism carries a 'small-scale model' of external reality and of its own possible actions within its head, it is able to try out various alternatives, conclude which is the best of them, react to future situations before they arise, utilize the knowledge of past events in dealing with the present and the future, and in every way to react in a much fuller, safe, and more competent manner to the emergencies which face it."

For our purposes, it is crucial that the embodied simulation formulation of the brain supports detailed theory and experiment on the neural realization of simulations. From my computer science and robotics background, simulation was always the obvious approach. Simulation seemed clearly basic in people (90) , but no one knew if other animals used it. Around 2006, I was able to start a research project on simulation in primates involving Jose Carmena at Berkeley and Mike Tomasello and Josip Call at the Max Planck Institute in Leipzig. The idea was to see if the animals could predict the outcome of a Pachinko pinball setup. Everything was more difficult than expected and the effort was abandoned.

Meanwhile, there have been remarkable advances in understanding the functions of the vertebrate hippocampal complex in memory and simulation. Various versions of these phenomena are studied as simulation (8), replay (56) and time travel (57). Over the past decades, some profound discoveries have come from studying *replay* in mammals.

" Hippocampal replay is a phenomenon observed in rats, mice, cats, rabbits, songbirds and monkeys. During sleep or awake rest, replay refers to the re-occurrence of a sequence of cell activations that also occurred during activity."     Wikipedia 6/1/20.

The main research focus has been the learned and exploited activity sequences in the hippocampus of rats. There are now several variants of both sleeping and awake simulation and replay in both the direction of motion and the reverse direction towards the starting point (9). These are characterized as high (theta) frequency oscillations, "ripples", in recordings of cells in the Hippocampus. They have been shown to play a significant role in memory and planning (58). In addition, the rate of ripples was significantly increased at rewarding versus non-rewarding sites (59).

This research was supported by similar findings in rats performing a goal-directed navigational task in an open arena (60). In this study, rats alternated between two behaviors in the same familiar environment: random foraging and goal-directed navigation to a recently learned location to obtain a predictable reward. The predictable goal changed location daily and goal-directed navigation to this newly learned position entailed unique combinations of start and end points. When the rat was away from the recently learned goal location, replay was strongly biased to encode spatial trajectories that started at the rat's current location and ended at the goal. Indeed, during goal-directed navigation, the rat's future behavioral trajectories were strongly correlated with the paths encoded by neural replay events. *It was unclear how the same replay mechanism could result in learning both forward and reverse trajectories. People do not automatically learn return routes when traveling.*

*More recently, understanding of the replay mechanism has significantly advanced. Remarkable new findings have (91) have suggested a paradigm that could be the basis of a wide range of simulation phenomena.*

*"These findings reveal an unexpected aspect of theta-based hippocampal encoding and provide a biological mechanism to support the expression of reverse-ordered sequences. … It remains unclear how forward-ordered neural activity can facilitate storage or expression of reverse-ordered sequences, which are observed in ripple-based reverse replay … However, theta oscillations rarely progressed uniformly in a single direction. They consisted of two distinct components, one that traveled ahead of the rat and a second that moved backward in the reverse direction of the animal's actual movement across place cells>"*

*Wang et al, (91) have conducted an ambitious experimental and computational analysis of rats simultaneously learning forward and reverse trajectories, extending previous work on theta rhythms in hippocampus place cells. Using a large sample of recordings ( >100,000 theta oscillations), they found that some cycles had two distinct firing patterns across the place cells.:*

*" theta oscillations rarely progressed uniformly in a single direction. They consisted of two distinct components, one that traveled ahead of the rat and a second that moved backward in the reverse direction of the animal's actual movement (Fig. 1, A and B, and figs. S1 and S2). The reverse component consistently occurred near the peak of theta oscillation (Fig. 1B); a time window associated with minimal hippocampal population activity."*

*They then set out to discover if the two patterns were correlated or came from different sources.*

*" the relative timing and independent expression that we observe in the forward and reverse components of theta sequences indicate that CA3 input selectively facilitates phase precession and drives the onward, prospective sequence, whereas antiphase entorhinal EC3 input selectively facilitates phase procession and drives the reverse, retrospective sequence. "*

*In addition, they suggest detailed cellular models of these circuits. Summarizing:*

*"we have specifically identified two distinct and independent information streams within each theta cycle: one that represents a possible future outcome and one that represents prior behavior in the reverse order. "*

*Wang et.al explicitly do not make such a claim (91), but the subtle phase interactions in their demonstration could be a paradigm example of complex neural computation.*

Further evidence that replay may provide a foundation for mental exploration of possible future actions comes from a study testing the role of replay in avoidance behavior ( 52).

After initial exploration of a linear track, rats were given a pair of mild shocks when they reached one end of the track. During subsequent exploration of the track, rats displayed consistent avoidance of the shock zone, stopping and turning around before entering it. Neural replay recordings during these pauses reliably encoded trajectories leading into the shock zone immediately prior to the rat turning around, consistent with a model of replay as a memory retrieval system capable of providing outcome predictions. Importantly, in this study the replay denoted paths to avoid rather than paths to follow, indicating that the content of replay is used to inform rather than dictate future behavior, possibly by coordinating the reactivation of amygdala-based representations of the valence of the encoded experience.

In summary, the rat hippocampal complex recalls and exploits a neural trace of actions and their valence. These records are used for memory consolidation in sleep and for planning and simulation when awake. As discussed above, spatial locations of positive and negative valence are recorded and used to modify behavior. This is a powerful illustration of the neural basis of emotions. The central issue of the postulated Chi mapping from neural to mental activity remains mysterious, but the rich replay and simulation literature is one of the best topics to explore for the phenomenon of human SE.

Some recent results suggest that songbirds have a parallel song *replay* mechanism (61).

"Recently, songbirds were found to have surprisingly mammalian-like sleep architecture as well as sleep replay of song, a learned motor skill. In mammals, sleep replay in hippocampal place cells is thought to be a central mechanism for the function of sleep in learning and memory. The well-established co-occurrence of hippocampal replay with slow waves demonstrates the link between global sleep architecture and circuit-level mediators of sleep function. However, sleep has profound effects not only on hippocampal-dependent declarative memory, but also on procedural memory. These results establish for the first time similarities and differences in sleep replay comparing declarative and procedural memories. In sum these results show that highly complex sleep traits manifest across songbirds and parrots, and that complex sleep architecture is linked to song replay. This supports the hypothesis that shared attributes of avian and mammalian sleep are derived from a common precursor, and helps to illuminate underlying mechanisms by which complex sleep can affect procedural memory. "

The conclusion so far is that any close evolutionary precursors to human SE are most likely to be found in advanced vertebrates. Mammals are particularly relevant, but the fact that (some) birds exhibit related SE suggests that an important component of human SE appears in the general amniote brain architecture, especially in the brain stem and related structures. There is considerable converging evidence (26) that these sub-cortical structures continue to play an important role in human SE and consciousness.

More generally, dreams, simulation and replay exhibit *self-generated activi*ty that could be a basis or precursor of subjective experience, as will be mentioned in the Conclusion section. Of course, people have a much wider range of mental experience including language, self-awareness, and Meta-cognition, which were labeled C2 in the Preface. Since the current article focuses on evolutionary precursors of human SE, all of this magic is beyond its scope. However, primates who lack language and probably C2 , are being productively studied to explore advanced aspects of SE. This is the topic of the next section, Primates.

## 4. Primates and Embodiment

We have seen that birds and mammals each exhibit a wide range of behaviors that are similar to those of human Subjective Experience (SE). Their shared precursors, the earliest vertebrates called amniotes, exhibit brain stem structures and activities that are also involved in human SE and consciousness (26). And, of course, human subjective experience is the ultimate target of this article. This section focuses on Primates who, as research subjects, have more human-like behavior than other vertebrates and also have brain anatomy and functionality more closely related to human SE.

There are two important benefits to studying SE in non-human primates. The first obvious advantage is that it is permissible to perform invasive experiments on primates, although fortunately much more constrained than in the past. We will examine multiple studies that localize *individual neurons* that embody subjective decisions or can be trained to adapt to body extensions like the cane of a blind person. The other advantage is that primates have richer internal models (17) and can be trained to carry out more complex acts of perception and action than other vertebrates. This functionality supports much richer subjective reports of experience, although still well short of human responses such as language.

We will concentrate on two remarkable projects on the detailed neural basis for primate phenomena in vision and action. These are related to the powerful SE found in people. Some of the core examples of human SE in perception involve ambiguous images such as the famous face-vase image (Figure 5), which can be experienced as two different scenes. The mechanism underlying this phenomenon has been the subject of considerable effort, especially by Rudiger van der Heydt and his colleagues. In addition to a wide range of behavioral experiments (63), they have been able to discover individual neurons in monkeys that play a crucial role figure-ground separation and related ambiguous perception.

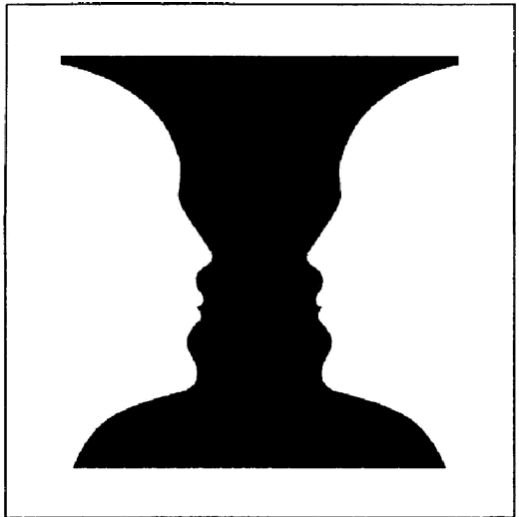
Figure 5. Classical Rubin (1912) Face-Vase image

Figure 5 presents a paradigm example of the experimental power of primate studies. Behavioral experiments had shown that the resolution of an ambiguous image depends on the assignment of border elements to either the inside or the outside of the perceived figure. A border element is essentially always "owned" by the figure and not the ground. In the Figure above, the small white triangles that are the nose of the face percepts become part of the background for the vase percept.

Building on this long history of behavioral studies, the von der Heydt lab set out to find direct neural correlates of border ownership in macaque early visual cortex, areas V1 and V2 (63). One key finding was that, for each border position, there is a complementary pair of neurons one for figure ownership of the edge and the other for the edge being part of the ground. The original study was for static images of overlapping squares and it has since been extended in several ways (64). Even the initial research has had a significant impact on the field because it was the first demonstration of a direct neural substrate of a famous SE shared by monkeys and people. This has led to continuing efforts to find neural correlates of other SE in perception and other experiences.

One subsequent project has explored the important role of border ownership in the recognition of proto-objects. Another established that the same neural results hold for images of natural scenes (64). The van der Heydt team has more recently specified a detailed computational model (65) that involves top-down feedback from higher visual areas that aggregate mutually consistent border cells into groups. Such neural groups are persistent, interact with attention, and map over small changes in the image or eye position. This interaction of bottom up and top down processing is a cornerstone of modern internal brain models (17).

**Embodiment**

More generally, the continuing effort to specify neural substrates highlights the unique possibilities of primate studies of SE. Finding specific neural realization of phenomena like Figure 5 is a paradigm example of *embodiment* in the mind-body problem. People, experience SE with these and many other activities including listening, reading, empathy, imagining, etc. and these also presumably have analogous neural realization. Further consideration of the SE of human visual perception, its problems, and possible therapies will be presented in Section 5. People.

So far, this article has continued the tradition of focusing on SE in the domains of emotion and perception, with some side discussion of dreams in non-human animals. In fact, subjective experience is important in many other aspects of mental life. We will now focus on a largely separate domain, motor control. One central question concerns the subjective embodiment of tools, prototypically the cane used by blind people (68).

Tool embodiment is not normally considered in general discussions of SE. One possible reason is that much of the emphasis in this field is on prosthetics and other tools of clinical importance (69). Prosthetics will be discussed in the following section, People, and in the Conclusion. However, tool embodiment provides some of the most scientifically interesting and challenging examples of SE, especially in monkeys.

The general phenomenon is that, after significant experience, the tip of a tool (like the blind man's cane) is experienced as a body part and is directly controlled. This SE is quite common and you may well have experienced it yourself with a musical instrument or a hand tool like a screw-diver. There is even anecdotal evidence the operators of a heavy equipment like a giant crane come to experience the tip of the crane as a bodily extension.

For many years, tool embodiment was treated as a deep mystery of SE, with no plausible solution. The whole idea of an external tool being incorporated directly in the body's sensing and control systems was counter-intuitive. The initial breakthrough and much of the subsequent progress was the work on monkeys by Atusushi Iriki and his collaborators (66).

In the initial 1996 experiment, they trained macaque monkeys to retrieve distant food using a rake, and recorded neuronal activity in the caudal postcentral gyrus where the somatosensory and visual signals converge. There they found a large number of bimodal neurons, which appeared to code the body schema of the hand. After extensive tool use, the visual receptive fields of these cells were altered to include the entire length of the rake or to cover the expanded accessible space. This effect did not appear when the monkey held the rake, but did not use it.

Additional studies revealed that after the tool-use training, monkeys began combining tools to reach even farther, a skill they were never taught. And the researchers discovered that these behavioral changes were accompanied by changes in gene expression in the parietal cortex and by increased intracortical connections between the temporo-parietal junction and parietal cortex (70). This adds important detail to understanding the mechanism underlying embodiment.

In his 2019 Keynote Address at the International Convention of Psychological Science, Iriki discussed his research-based insights into the evolutionary phases of brain development and hypotheses about what may be coming next.
https://www.psychologicalscience.org/observer/iriki-keynote

In summary, after years of trial and error, Iriki and his team were able to expand the monkeys' tool-use repertoire to include tools that augmented their sensory ability. The researchers took a hand rake, similar to those used in the earlier studies, and attached a big mirror at the end, adding a sensory cue to a tool that was already integrated with the monkeys' body image. The monkeys were then able to explore the surrounding environment with the mirror and reach for the food with the rake. Several training steps later, the monkeys were able to use a small camera, similar to an endoscope, attached to the end of the rake.

In more recent work (70) Iriki has pursued the examination of tool embodiment as a core aspect of evolution.

"Hominin evolution has involved a continuous process of addition of new kinds of cognitive capacity, including those relating to manufacture and use of tools and to the establishment of linguistic faculties. The dramatic expansion of the brain that accompanied additions of new functional areas would have supported such continuous evolution. Extended brain functions would have driven rapid and drastic changes in the hominin ecological niche, which in turn demanded further brain resources to adapt to it. In this way, humans have constructed a novel niche in each of the ecological, cognitive and neural domains, whose interactions accelerated their individual evolution through a process called *triadic niche construction* "

At this point, the embodiment of tool use has become "An open window into body representation and its plasticity". The recent paper (67) of that name is a critical review of the way different tool-use paradigms have been, and should be, used to try to disentangle the critical features that are responsible for human tool incorporation into different body representations, an example of internal models. This study focuses on the role of tool use in human peri-personal space and the effects of neurological damage or Trans-cortical TMS intervention. It reviews a wide range of experimental paradigms and examines several open questions.

There are two important scientific conclusions from this line of work. The more obvious one is the fact that growth of animal bodies requires neural adaptation so that the body image of peri-personal space matches the changing size of the body. Therefore, normal development requires some adaptation (70). The continuing insightful research on primate tool embodiment goes far to explain the neural substrate, but does not show whether the primates have the SE. The other insight, forward models (17) , is more complex and could play a key role in the understanding of human SE. I will discuss this in some detail in the next section, People.

## 5. People

Human subjective experience was the original source of the mind-body problem. This mystery was discussed in various ways by the Greeks (71**)** https://plato.stanford.edu/entries/ancient-soul/ and also in other contemporary cultures. The basis for the modern formulation is generally attributed to Descartes (72). One reason for the continuing importance of the mind-body problem is the presence of ubiquitous mysteries of everyday experience (11).

And, of course, human language introduces a wide range of additional SE issues. We will not address these here, but our approach to language also relies heavily on embodiment and simulation (8, 81). Research on the evolution of language is making rapid progress, the recent excellent survey (80) is a good place to start.

An intriguing recent paper from the Iriki group (70) suggests how embodied tool use could be a crucial step in the evolution of language.
 "Humans have constructed a novel niche in each of the *ecological*, *cognitive* and *neural* domains, whose interactions accelerated their individual evolution through a process of *triadic niche construction*. Human higher cognitive activity can therefore be viewed holistically as one component in a terrestrial ecosystem. The brain mechanisms that subserve tool use may bridge the gap between gesture and language - the site of such integration seems to be the parietal and extending opercular cortices."

The first section of the current paper, Prologue, contains an extensive discussion of human Subjective Experience and its relation to the general mind-body problem. In the present section, the focus will be on some more technical results and how they add insight to our understanding of the mind-body problem.

Some profound results on the neural basis of SE have been developed in the context of human behavior.  The 1942 quote from Craik (55)  cited in the previous section is a famous precursor to current work on internal models and simulation.  For our purposes, some of the most important work concerns "Internal Models in Biological Control" summarized in (17). The general notion of "internal models" has been a cornerstone of the current article. All organisms and other agents must include internal mechanisms that relate inputs to actions and these have now been explored in considerable detail.

As is well known, many human motor actions (e.g. playing a piano) can be too fast to depend on sensory feedback and thus must rely on internal models. Some of the most elegant and effective work on human internal models has been on the control of complex arm actions. These involve deep formalisms from engineering including Bayesian (probabilistic) formalisms and sensorimotor feedback control. In formal sensorimotor control, the specification of a particular behavioral task begins with a definition of what constitutes the relevant internal state *x* (which may include components corresponding to the state of the arm and external environment) and control signals *u*. Of course, it is the *constancy* values (e.g. size, shape) of the external world state that are needed for achieving effective action.

In general, the state variables should include all the variables which, together with the equations of motion describing the system dynamics and the motor commands, are sufficient to predict future configurations (in the absence of noise). For reaching movements, the state *x* could correspond to the hand position, joint angles, and angular velocities, and the control signals *u* might correspond to desired joint torques.

I will introduce these ideas following the excellent recent review by McNamee and Wolpert (17), using their example of tennis. Consider the problem of tracking a ball during a game of tennis. The response of any given photoreceptor in a player's retina can provide only delayed, noisy signals regarding the position *y* of the ball at a given time. An internal model is needed to *simulate* the probable trajectories of the ball. This internal forward dynamical model must take into account physical laws, such as air resistance and gravity. From a perceptual point of view, new sensory information *at* time t+1 can then be integrated with this predictive distribution in order to compute a new posterior distribution of positions at time *t* + 1.

This iterative algorithm, known as Bayesian filtering, can be used to track states of the body or the environment in the presence of noisy and delayed signals for the purposes of state estimation. The results of such computations are advantageous to the tennis player. On a short timescale, they enable the player to predictively track the ball with pursuit eye movements, while on a longer timescale, the player can plan to move into position well in advance of the ball's arrival in order to prepare the next shot.

In the sensorimotor context, internal models are broadly defined as neural systems that mimic musculoskeletal or environmental dynamical processes. An important feature of putative internal models in sensorimotor control is their *dynamical* nature. This dynamical nature is reflected in the brain computations associated with *active* internal models.

Internal models that represent future states of a process (e.g., ball trajectories) given motor inputs (racquet swing) are known as *forward* models. Conversely, models that compute motor outputs (the best racquet swing) given the desired state of the system at a future time point (a point-winning shot) are known as *inverse* models. For human motor expertise, both kinds of models must be learned. Basic Bayesian feedback models have been used successfully in reinforcement learning for a wide range of robotic tasks (17).

Whether contributing to state estimation, reafference cancellation, or planning, internal forward and inverse models relate world states across a range of temporal scales. In the tennis example above, internal models may be used to make anticipatory eye movements in order to overcome sensory delays in tracking the ball. By incorporating a motor response, forward internal models can be used to *simulate* the ballistic trajectory of a tennis ball after it has been hit.

*Forward models* play a central in our consideration of the SE of *embodiment*. Recall the discussion of the Iriki research on tool embodiment (66) in Section 4, Primates. After considerable training with a tool like a rake, the tool was incorporated into the monkey's internal model of itself, as shown by changes in neural receptive fields. However, we can also view the monkey's internal model from the perspective of optimal control theory (17). In developing the skill with a new tool, the monkey needs to learn a forward model of the tool use in the required tasks. According to my Chi hypothesis, Chi then maps the embodied forward model to SE, the same as for natural neural activity. There is no evidence that the monkey's internal model is anything like the Bayesian theory.

In people, there is a similar phenomenon of tool embodiment, prototypically the blind man's cane, which comes to feel like an extended body part. Crucially, tool embodiment in people entails the Holy Grail of reported subjective experience, SE. Moreover, reported SE is now known to be a fundamental criterion for the success of prostheses (74).

The McNamee and Wolpert paper (17) also reviews the neural computational roles played by such internal models and the clinical and behavioral evidence for their implementation in the brain. In addition, there is thoughtful discussion of the technical problems that arise in more complex real-world problems of control and planning. We will further explore motor control and its embodiment as prostheses at the end of the Conclusions section.

**Blindsight**

One interesting class of human SE issues involves the inconsistency of visual description with visually guided action. One famous case has been called "blindsight" https://en.wikipedia.org/wiki/Blindsight. People with certain deficits will not be able to report what they see but can carry out appropriate actions like grasping a tool. Until fairly recently, this has been viewed as a deep mystery. See, for example, this BBC program: https://www.bbc.com/future/article/20150925-blindsight-the-strangest-form-of-consciousness

In retrospect, the mystery of blindsight arose from the simplistic assumption that visual perception was a single integrated function. For our main concern of SE, a particularly important recent development is the widespread clinical and experimental work on the "two streams" model of vision: https://en.wikipedia.org/wiki/Two-streams_hypothesis. This postulates that the ventral visual stream leads to the temporal lobe, which carries out object identification and recognition. The dorsal stream leads to the parietal lobe, which is involved with processing spatial location relative to the viewer and with motor responses. The basic model has been very productive in clinical and experimental research, but it is now clear that it is not the whole story.

"We should view the model not as a formal hypothesis, but as a set of heuristics to guide experiment and theory. The differing informational requirements of visual recognition and action guidance still offer a compelling explanation for the broad relative specializations of dorsal and ventral streams. However, to progress the field, we may need to abandon the idea that these streams work largely independently of one other, and to address the dynamic details of how the many visual brain areas arrange themselves from task to task into novel functional networks" (71).

In addition to the various cortical interactions, there are sub-cortical systems that can "hijack" the response to (visual) input. https://en.wikipedia.org/wiki/Amygdala_hijack

" The output of sense organs is first received by the thalamus. Part of the thalamus' stimuli goes directly to the amygdala or "emotional/irrational brain", while other parts are sent to the neocortex . If the amygdala perceives a match to the stimulus, , then the amygdala triggers the HPA (hypothalmic-pituitary-adrenal) axis and hijacks the rational brain. This emotional brain activity processes information milliseconds earlier than the rational brain, so in case of a match, the amygdala acts before any possible direction from the neocortex can be received. When the amygdala perceives a threat, it can lead that person to react irrationally and destructively."

Taken together, these results reveal that the oculomotor system has access to motion information that is, at least in part, distinct from information used to support conscious motion perception. In an exclusively subcortical route, the interactions between the Superior Colliculus (SC), pulvinar, and the amygdala have been considered essential for the mediation of affective blindsight in both humans [45] and monkeys [46]. Some cases involve individuals being able to correctly discriminate emotional stimuli presented within their blind field; for example, distinguishing between happy and fearful faces that are presented in a forced-choice manner [47]. These findings support the notion of an unconscious visual pathway that can extract affective features from facial expressions without input from higher-order areas of the ventral visual stream, involved in face and object recognition, as well as to the absence of V1 input.

Due to the presence of a broad range of preserved behavioral capacities in blindsight patients, there has been a shift away from identifying a unitary neural substrate of vision. Instead, blindsight is thought to operate through multiple functional groups, with each group preserving an element of unconscious visual function, such as the sensitivity to motion direction. This is a paradigm example of a recent *demystification* in the mind-body problem, which will be revisited in the Conclusion.

## 6. Conclusions

*In formal logic, a contradiction is the signal of defeat, but in the evolution of real knowledge it marks the first step in progress toward a victory.* **Alfred North Whitehead**

The ancient *mind-body problem* continues to be one of deepest mysteries of science and of the human spirit. Despite major advances in many relevant fields, there is still no plausible causal link between human subjective experience (SE) and its possible realization in the body. This paper suggests that considerable progress is being made on scientific constraints on any plausible theory of SE and points out promising avenues of research. Independent of the deep realization mystery, we can explore how SE became a *stable trait* and evolved to be the hallmark of the human mind. In addition, several SE phenomena are known to be incompatible with contemporary neuroscience so some new results and insights would be needed to bring them into current Science.

Often in the history of Science, such anomalies have provided leverage for demystifying gaps in our understanding.

As scientists, we must acknowledge that there is a world external to our minds and that we have no privileged access to it. Formal systems are the best vehicles for expressing theories about the world, but
1) Any adequate formal system cannot prove itself complete and consistent.
2) A formal system or systems can be proven inconsistent.
3) "     "         "            "      can be shown to be empirically wrong.
4) Science proceeds by testing and modifying formal descriptions.
5) Modern computation greatly facilitates testing and modifying theories.

The historic Enlightenment project was based on the false assumption that everything is understandable by the human brain/mind. At any time, there are profound mysteries of the material, spiritual, and social world. Science pursues demystification, but should remain agnostic about ultimate success. An insightful analysis of the Enlightenment project as a basis for society can be found in
https://philosophynow.org/issues/79/Whats_Wrong_With_The_Enlightenment

One core issue in the mind-body problem is the distinction between routine $3^{rd}$ person science and the agent based ($1^{st}$ person) experience. The mind-body problem is a deep mystery because it involves both internal and external perspectives that have no known common ground. Many famous problems of every day perception are incompatible with current theories of neural computation(11) . Evolutionary success is based largely on current fitness of the phenotype, but an organism can only make an internal estimate, *actionability*, of the fitness of its actions (8). Even very simple organisms can be studied from an internal agent perspective and some old principles such autopoesis and umwelten have proven valuable in our search for the evolution of SE.

A wide range of converging evidence supports the notion that the vertebrate branch, which includes people, mammals, and birds, contains the most developed SE. More specifically, the vertebrate brain stem appears to be necessary for global mental states like sleep and SE. This structure is largely preserved and is essential in advanced vertebrates, including people. This can be viewed as a plausible lower bound for SE, as discussed in the Vertebrates section.

Our working hypothesis is that at least a primitive SE is active in (many) birds and mammals. There are two main reasons for this suggestion. For one thing, no one seriously argues that e.g., dogs, do not experience emotions in ways that seem human like. Also, more technically, some mammals and birds are known to employ mental simulation, including dreams, in planning and learning. So, we should expect some evolutionary change, probably involving simulation, at the vertebrate level. In fact, recent studies (9, 60) have provided extensive description of simulation involving the hippocampus of rodents.

However, there remains a large gap between the minds of other mammals and those of people. An intermediate stage can be found in primates. Primates can be trained to produce complex responses and behaviors and also permit more invasive experiments. Some human mysteries, like perceptual puzzles and tool embodiment have been elucidated by single cell experiments in primates (63-65).

The SE of people is our main concern. Much of the basis of human SE can be found in mammals and primates. However, people alone can directly compare their experience with scientific theories and experiments. Several of the results discussed in this paper show promise of helping to understand the human mind. One important advance is the recognition that a human ability, such as vision, is not a single system but a complex of interacting cortical and subcortical networks. This has led to the understanding of phenomena like Blindsight (76), which had been considered a deep mystery.

**Science as Demystification**

A core mission of Science is attempting to explain the mysteries of nature. The history of science is largely a saga of increasingly sound theories of the physical and social world. There is broad agreement that the nature of the mind is one of the deepest current mysteries and one might hope that science will help demystify it. For now, the only plausible scientific stance is *agnostic mysterianism* (30) – acceptance of current mysteries and research on the boundaries between the known and the unknown, without assuming ultimate victory. Much of the historic success of Science has followed this paradigm and there currently are promising relevant efforts on demystification of the mind.

A *scientific problem* or *mystery* is a phenomenon for which there is currently no plausible explanation. A related source of mystery is an *inconsistency* between two or more theories of the same phenomena. A previous article (11) made the case that the mind-body-world problem is *inconsistent* with current neuroscience and computational theory. Such inconsistencies often lead to scientific revolutions. Much of the historical success of science can be traced to concerted effort on mysteries. The evolutionary analysis in this article provides some useful constraints and suggestions for exploring subjective experience.

This paper largely follows the current assumption that SE and Consciousness (Cs.) will eventually be explained scientifically and explores the constraints introduced by evolution. In fact, I believe that there is an unbridged conceptual gap between existing science and the deep problems of the mind (30). This article focuses on one particular mystery of the mind, subjective experience, but there are many others. I explicitly deferred consideration of reflection (C2, 6) as well as body image, self, free will, learning, memory, development, and social phenomena. Any science of the mind will need to incorporate all of this and more. My hope is that the present analysis will suggest conceptualizations and experiments that lead to some further demystification.

Perhaps the most significan*t* scientific demystification ever was the eventual disproof of claims for a life force or *Vitalism.* This is the belief that "living organisms are fundamentally different from non-living entities because they contain some non-physical element or are governed by different principles than are inanimate things"

The Vitalism theory was challenged in 1852 by Friedrich Wohler, who showed that heating silver cyanate (an inorganic compound) with ammonium chloride (another inorganic compound) produced the organic compound urea, without the aid of a living organism or part of a living organism. This was not a definitive proof and the controversy lasted for several decades, but now Vitalism is a superseded scientific hypothesis, and the term is sometimes used as a pejorative epithet.   However, [Ernst Mayr](#) (79) wrote:

"It would be ahistorical to ridicule vitalists. When one reads the writings of one of the leading vitalists like Driesch, one is forced to agree with him that many of the basic problems of biology simply cannot be solved by a philosophy as that of Descartes, in which the organism is simply considered a machine... The logic of the critique of the vitalists was impeccable. In fact, the origin of life is still unsettled and analysis is parallel to the current discussion of dualism and reductionism in the mind-brain problem."

Several other ancient mysteries of the mind have been largely reduced to routine science within our lifetime. One interesting case is synesthesia (15), a perceptual experience in which stimuli presented through one modality spontaneously evoke sensations in an unrelated modality. This is discussed in (30). Another current demystification is underway in the evolution of human language (80). This advance is based on reframing the problem as the co-evolution of genome and culture in the eco-evo-devo research program [https://en.wikipedia.org/wiki/Evolutionary_developmental_biology](https://en.wikipedia.org/wiki/Evolutionary_developmental_biology) .

 One striking recent demystification is the profound work of the von der Heydt group, reviewed section 4, Primates. For many years, visual illusions like the famous face-vase image of Figure 5 remained mysterious. In a continuing series of experiments and models, von der Heydt (63-65) has illuminated a neural and computational substrate of these phenomena. This is a significant advance in the naturalization of SE/qualia.

Another important current example of demystification involves the phenomenon of Blindsight (76) discussed in the People section. It long seemed mysterious that people with certain visual deficits could perform tasks like visual navigation and tool use while being unable to describe their visual experience. It is now understood that vision (as well as other traits) is embodied by a complex of interacting circuits. A remarkable consequence of this new knowledge is the development of diagnoses and therapies for some visual deficits; this will be discussed as a clinical application of SE research later in this section.

**How SE evolved as a stable trait.**

Evolution selects on whatever works. We have discussed many examples where fitness is improved by non-veridical representation of the internal and external world. For our Origins examples, this fitness constraint is nicely captured by the Umwelt idea, discussed in Section 2. For SE questions in vertebrates (e.g. perceptual constancies), we know that complex models and computations are involved.

Evolutionary change involves modifications of the genotype that enhance fitness, but this can happen at many different levels. For example, color constancy of the Hawkmoth evolved through modification of the photoreceptor molecule to selectively respond to changes in the ambient spectrum. Color and other adaptive constancies evolved through a wide range of lower and higher mechanisms, which are not veridical and can be considered internal models. Such evolutionary fitness mechanisms can be seen as precursors of SE.

A useful generalization is to call these adaptations "Internal Models". There is a developing field studying internal models (17) and we have exploited this idea in our treatment of subjective perception and tool embodiment. *From the current perspective, the central issues are how internal models and simulation are represented in the brain and how they are experienced in ways that lead to adaptive behavior.*

When it comes to evaluating the possible outcomes of various alternatives, an animal needs to evaluate the possible resulting subjective feelings. As mentioned above, this was realized in 1943 by Craik(55). The crucial fact is expressed explicitly in the following quotation from (78):

"A common way to evaluate outcomes is to anticipate how we would *feel* if they happened, and this has been called *affective forecasting* " (emphasis on *feel* in original). This same ability to anticipate future emotions and organize current behavior accordingly underlies flexible and advanced preparation for future dangers "

Converging evidence indicates that birds and mammals have rudimentary SE and also effective mental simulation (often called *replay*), described in some detail in the Vertebrates section. This shows that, even in rodents, internal models and simulation are highly adaptive in behavior, learning, and memory.

The field called "affective forecasting" is a richly evolving discipline with a wide range of applications https://en.wikipedia.org/wiki/Affective_forecasting, but the central idea is that SE feelings arising from *simulation* of possible futures are essential for planning. All of this has now been made much more precise in the Bayesian Optimal Feedback theory, which was discussed in the People section. Simulation is now viewed as the core computational mechanism for remembering or imagining the past and modeling the future (82, 83), all of which require SE. More generally, Kanai et al. (86) have developed a case for simulation as the foundation of Consciousness, which obviously also entails SE.

In the current context, this says that the SE, which are known to be present in mammals (and birds), are the only known way that a creature can evaluate the subjective valence of a simulated plan. Therefore, given the SE in mammals, it was adaptive to link this trait to planning based on simulation involving internal models of the physical and social world. There is good evidence that primates do sophisticated social planning (31) as well as the remarkable findings of Iriki on the embodiment of tools by primates, as discussed in the Primates section. Following my Chi hypothesis, when a monkey builds an internal forward model of a tool, this causes him to experience the tool as a body part.

*From my 2021 perspective, it is useful to consider Subjective Experience(SE) from the perspective of Tinbergen's famous four criteria for traits [https://en.wikipedia.org/wiki/Tinbergen%2s_four_questions](https://en.wikipedia.org/wiki/Tinbergen%2s_four_questions)) : (1) adaptive function , (2) phylogenetic history, (3)underlying physiological mechanisms and (4) ontogenetic/developmental history. There are many other cases of SE where your experience is inconsistent with current neuroscience. For example, if you touch your nose with your thumb, it is experienced as simultaneous despite the large difference in neural conduction times (also cf. 95). The adaptive function (1) of such SE is to provide a useful internal substrate for action and planning. The evolutionary(2) and developmental (4) aspects of SE are discussed in detail in this paper. It is only the realization (3) and the mapping Chi to SE that remain mysterious.*

So, SE is an essential part of advanced planning and should be selected for in creatures like ourselves. A further insight comes from Van Boven and Caruso and their colleagues who have shown that people experience more intense emotions when they anticipate future experiences than when they retrospect about either actual or hypothetical past experiences (84).

Of course, the realization of all this in the body remains part of the hard problem. However, existing insights into bodily correlates of SE are already being used to develop clinical therapies and prostheses.

**Clinical Applications**

Emotional and other subjective experiences play an essential role in human life including in physical and mental health. For our focus on the science of subjective experience, there are some particularly relevant efforts on therapies and prostheses. Building on the technical descriptions in the previous section, People, I will discuss some promising links between scientific research on SE and applications in health care.

Blindsight long seemed to be a mysterious ability of some blind people to successfully perform tasks like tool manipulation and spatial navigation. The confusion arose from a simplistic view of vision as a unified capability; this view persisted well into the 20[th] Century. Some relevant developments are outlined in the People section above. The key finding was that perceptual awareness and linguistic description involved the ventral pathway, but many other visual functions did not. The spared visual abilities observed in blindsight vary across a broad range of behaviors, including visually guided actions, such as pointing towards or grasping an object and navigating obstacles; discriminating among emotional signals expressed on faces as well as a number of cognitive processes such as attention and spatial memory. *In fact, many people have had the experience of physically reacting to a threatening scene before being aware of its content.*

More recently, people have begun to investigate these alternative visual pathways as pathways for aiding or training abilities of subjects with deficits. The state of this effort as of 2020 is beautifully summarized in the review "The Age-Dependent Neural Substrates of Blindsight" (76). As the title suggests, various blindsight problems are now known to rely on different neural substrates, depending in part on when the problem arose. This is an instance of the common phenomenon that trauma that occurs earlier in development is often treatable or even internally compensated.

The well-known ventral visual recognition stream goes through a specialized structure of the thalamus called the lateral geniculate and then to primary cortex V1 and beyond. More recently, there is strong evidence that an alternative thalamic structure, the pulvinar, provides a parallel pathway beyond V1 to area MT, especially early in development.  It has been argued that some of these pathways, particularly those engaging the cerebral cortex, may have been rewired after damage to V1, mainly in cases where the damage occurred early in life (76).

Advances in functional and connectivity imaging have revealed significant insights into the networks facilitating the preserved visual abilities seen in blindsight patients. Once it is better understood how the visual system responds to primary visual cortical injury, one could employ rehabilitative strategies to enhance (or even direct) this process with the ultimate aim of improving residual visual capacity. *Indeed, repeated exposure to visual stimuli within the blind field of V1 lesioned patients can yield to the rescuing of residual function, presumably through geniculo-extrastriate projections [87].*

*Another important and developing clinical application of embodied SE is teleoperated surgery. As we have seen in monkeys and humans, an external device (like the blind man's cane) can come to be experienced as an embodied extension. Unsurprisingly, the multiple remote instruments in tele-surgery can also come to be directly experienced. This aspect is not usually emphasized, but comes across strongly in journalistic presentations like this recent article from The New Yorker* [https://www.newyorker.com/magazine/2019/09/30/paging-dr-robot](https://www.newyorker.com/magazine/2019/09/30/paging-dr-robot) .

"A user as skillful as Dr. Giulianotti creates the illusion of having three operative hands; surgeons who regularly use the da Vinci arm often report experiencing a heightened sense of control.

"I felt the small robotic hands of the robot were a prolongation of my own. If you are used to having flat vision, and you pass into 3-D, you feel you are *immersed* inside the human body. It was a fantastic journey—the interior of the anatomy, the shadow of little vessels and nerves. I immediately fell in love." "

**Prosthetics**

The role of human SE in prosthetics has also advanced and is an active and growing field (74,77). Augmenting human capabilities with artificial devices has an ancient history. The first confirmed use of a prosthetic device is from 950–710 BC. In 2000, research pathologists discovered a mummy from this period buried in the Egyptian necropolis near ancient Thebes that possessed an artificial big toe. This toe, consisting of wood and leather, exhibited evidence of use. When reproduced by bio-mechanical engineers in 2011, researchers discovered that this ancient prosthetic device enabled its wearer to walk both barefoot and in Egyptian style sandals (https://www.livescience.com/23642-prosthetic-toes-egypt.html). It seems likely that the wearer of this device came to experience it as part of his body.

The most common embodied prosthetic device is the cane used by people with compromised vision. The use of a cane by individuals with visual impairment dates back to antiquity; a short stout cane was traditionally tapped to warn others to get out of the way and to set up echoes as clues to the environment (68) . Although this is not usually stressed, the user of the cane comes to experience the tip of the cane directly as a body part. From our perspective, this is essentially related to the profound discoveries of tool embodiment by the Iriki group discussed in the Primates section. It is also linked to the many instances of tool embodiment in people, as discussed in the People section.

In addition to use of external tools like the cane, there is enormous interest in prosthetics that are directly linked to the body. These can include visual and auditory and even dental implants, but the relevant research concerns on limb and body prosthesis. There is impressive and continuing development of artificial effectors, often overlapping with advances in robotics (74,77). Unfortunately, the state of sensory feedback systems lags far behind. While indirect visual feedback has some benefits, the Holy Grail is explicit *embodiment*, the direct incorporation of the prosthetic in the user's body image. This goal is now conventional wisdom in the field (74):

"We discuss potential clinical benefits of enhanced embodiment of the external objects by way of multisensory interventions. This review argues that the future evolution of human robotic technologies will require adopting an embodied approach, taking advantage of brain plasticity to allow bionic limbs to be mapped within the neural circuits of physically impaired individuals."

However, the difficulties of doing this are also understood (69):

"Invoking embodiment has shown to be of importance for the control of prosthesis and acceptance by the prosthetic wearers. It is a challenge to provide (conscious) feedback to cover the lost sensibility of a hand, not be overwhelming and confusing for the user, and to integrate technology within the constraint of a wearable prosthesis."

*The enormous benefits and challenges and of embodied prostheses are the subject of many efforts around the world. An excellent indication of the situation as of early 2020 is presented in (77). This pilot study compared various ways of achieving embodied prostheses against a range of tasks and subjective questionnaires. The paper includes detailed descriptions and illustrations of the anatomical arrangements and experimental findings. There are a number of supplementary figures, including several videos. They frame their experiment as follows:*

*"We implemented a hybrid approach for restoring multimodal sensory information to transradial amputees, where finger position information (referred to as remapped proprioception) was provided using sensory substitution based on peripheral intra-neural stimulation, whereas tactile information was restored using a somatotopic approach, where the elicited sensation was correctly perceived on the fingers and palm. Furthermore, because sensory substitution can be implemented using a large number of approaches, we performed a side-by-side comparison of our results with the same sensory substitution approach implemented using noninvasive electrotactile feedback with one participant. "*

*They then performed a range of detailed quantitative tests. One of these involved estimating the size of one of four cylinders. The two prosthesis subjects had an accuracy of 78% as compared with a 98.5% for five healthy subjects. The researchers also explored several variants on this size task, some of which also involved estimating compliance.*

*"Several control conditions were tested with participant 2. First, the same task was repeated with tactile feedback alone. In this scenario, performance was poor but remained above the 25% chance. However, further analysis showed that only the largest object was correctly identified above chance level. This indicated that tactile feedback alone was not sufficient to perform this task (i.e., recognizing all objects) "*

*"We tested whether using noninvasive electrotactile feedback as a source of sensory substitution would result in significantly different results compared with the same approach based on intraneural feedback. Specifically, proprioceptive acuity was lower when sensory substitution was provided using noninvasive stimulation compared with intraneural stimulation. "*

*They then carried out a series of studies of particular interest to our exploration of Subjective Experience.*

*"Last, we looked at prosthesis embodiment using a subjective questionnaire, which is a qualitative assessment and therefore intrinsically limited. However, because prosthesis embodiment is likely to play an important role in limb rejection rates, these types of questionnaires are increasingly used to evaluate sensory feedback approaches in prosthetics. In our case, it provides useful information when choosing between both strategies for the delivery of remapped proprioception.*

*Statistically significant differences between invasive and noninvasive sensory substitution were observed for certain questions but not for others, indicating that prosthesis embodiment was lower when providing remapped proprioceptive feedback using noninvasive electrical stimulation and higher when providing the same feedback using intraneural electrical stimulation. In both cases, embodiment was higher than in the absence of any stimulation.*

*Although both types of sensory substitution resulted in an improved sense of embodiment, superficial stimulation led to statistically lower answers on specific embodiment questions compared with invasive feedback. "*

*Unsurprisingly, although people can learn to embody feedback from remote body sites, it is more efficient and more natural to use feedback directly to the appropriate sensory*

*nerves. Although (77) is a specific study, the paper lays out and tests many of the central questions about embodied prostheses.*

**The End, for now.**

At the end, the ancient mind-body problem remains a mystery. There is no reason to postulate some non-material force that creates our subjective experience, SE. On the other hand, we know that existing neuroscience is inconsistent with this experience (11). But Science often advances as demystification and there are pertinent current examples, including tool embodiment and blindsight

The core problem is still the gap between 3rd person objective science and subjective, agent-centered experience. Considerable converging evidence suggests that SE is present in many vertebrates and that simulation (including dreaming) is present in mammals and birds. Internal models and simulation enable these animals to exploit past and future scenarios, but this requires something like internal SE for valence.

We also know that primates, including people, have a remarkable SE ability to embody tools as interacting body parts and that this can be extended to protheses. The most promising way forward is using the increasing drive for embodied therapies and prostheses to recharge a science of the mind.